\documentclass[a4paper,11pt]{article}

\usepackage{jheppub} 

\usepackage[T1]{fontenc} 
\allowdisplaybreaks[3]

\title{\boldmath Higgs phenomenon for higher spin fields on AdS$_3$}

\author[a]{Thomas Creutzig}
\author[b]{and Yasuaki Hikida}

\affiliation[a]{Department of Mathematical and Statistical Sciences, University of Alberta, \\ Edmonton, Alberta T6G 2G1, Canada}
\affiliation[b]{Department of Physics, Rikkyo University, \\ 3-34-1 Nishi-Ikebukuro, Toshima, Tokyo 171-8501, Japan}

\emailAdd{creutzig@ualberta.ca}
\emailAdd{hikida@rikkyo.ac.jp}

\abstract{In a previous work, a marginal deformation of 2d coset type model with $\mathcal{N}=3$ superconformal symmetry was studied, and it was interpreted as a change of boundary conditions for bulk fields in the dual higher spin theory. The deformation breaks generic higher spin gauge symmetry, and the generated mass of a spin 2 field was computed. The deformation might be related to the introduction of finite string tension in a superstring theory. In this paper, we extend the analysis and compute the masses of generic higher spin fields at the leading order of $1/c$ ($c$ is the CFT central charge) but at the full order of the deformation parameter. We find that the masses are not generated for so$(3)_R$ singlet higher spin fields at this order and the spectrum is the Regge-like one for so$(3)_R$ triplet higher spin-charged fields. 
}

\arxivnumber{1506.04465}

\begin{document}
\maketitle
\flushbottom

\section{Introduction}

There exist many higher spin states in superstring theory, and a higher spin symmetry is expected to appear at the limit where the masses of these states vanish. Moreover, it was proposed that the broken phase of higher spin gauge theory can describe superstring theory \cite{Gross:1988ue}. Recently it became possible to discuss the direct relation between higher spin gauge theory and superstring theory but on AdS space. This is due to the developments on higher spin gauge theory on AdS space such as Vasiliev theory \cite{Vasiliev:2003ev} and AdS/CFT correspondence. For examples, 4d Vasiliev theory is conjectured to be dual to 3d O$(N)$ vector model \cite{Klebanov:2002ja} (see also \cite{Sezgin:2002rt}), and a lower dimensional version was proposed in \cite{Gaberdiel:2010pz} where 3d Vasiliev theory in \cite{Prokushkin:1998bq} is dual to a 2d large $N$ minimal model.

The first concrete proposal on the relation between higher spin gauge theory and superstring theory was made in \cite{Chang:2012kt} by extending the duality in \cite{Klebanov:2002ja}.  
There are two other proposals given by generalizing the lower dimensional version of duality in \cite{Gaberdiel:2010pz}. Lower dimensional models are generically more tractable than higher dimensional ones, so we expect to learn more about the relation.
The proposal with large or small $\mathcal{N}=4$ supersymmetry was made in \cite{Gaberdiel:2013vva,Gaberdiel:2014cha,Gaberdiel:2015mra}, while that with $\mathcal{N}=3$ supersymmetry was given in \cite{Creutzig:2013tja,Creutzig:2014ula,Hikida:2015nfa}.
Compared to the proposal in \cite{Gaberdiel:2013vva,Gaberdiel:2014cha,Gaberdiel:2015mra},  the relation between higher spin fields and strings is more transparent in \cite{Creutzig:2013tja,Creutzig:2014ula,Hikida:2015nfa} like in the duality in \cite{Chang:2012kt}.
See \cite{Candu:2013fta,Gaberdiel:2014yla,Beccaria:2014jra,Candu:2014yva,Baggio:2015jxa} for related works.

Utilizing the features of the proposal in \cite{Creutzig:2014ula}, we have examined the Higgs phenomenon due to the breaking of higher spin symmetry in \cite{Hikida:2015nfa}\footnote{The Higgs phenomenon for the proposal in \cite{Gaberdiel:2014cha} was examined in a quite recent paper \cite{Gaberdiel:2015uca}.}.
 We deform the CFT such  that generic higher spin symmetry would be broken except for the $\mathcal{N}=3$ superconformal symmetry. The deformation is of the double-trace type and it can be interpreted as the change of boundary conditions for the bulk fields in the dual higher spin theory \cite{Witten:2001ua}. The breaking of higher spin gauge symmetry would make the higher spin fields massive. In \cite{Hikida:2015nfa}, we have computed the mass of a spin 2 field both from the bulk and the boundary theories by making use of the previous works  \cite{Porrati:2001db,Duff:2004wh,Kiritsis:2006hy,Aharony:2006hz}.
Similar phenomenon was also discussed for higher spin fields in \cite{Girardello:2002pp} but on AdS$_4$.

The aim of this paper is to extend the analysis in \cite{Hikida:2015nfa} to generic higher spin fields. We work at the leading order of $1/c$ with $c$ the CFT central charge, where the classical gravity computation is reliable. Utilizing the holographic duality, we succeeded to obtain the mass formula for all higher spin fields and at the full order of the deformation parameter. The bulk higher spin theory has $\mathcal{N}=3$ supersymmetry and there are so$(3)_R$ singlet fields with spin $s=2,3,4,\ldots$ and so$(3)_R$ triplet fields with spin $s=1,2,3,\ldots$. We observe that the masses are not  generated at this order for the so$(3)_R$ singlet fields. This result is actually consistent with that in \cite{Gaberdiel:2013jpa}, since their deformation is of the same type as ours. We also find  the mass square of spin $s$ field $M_{(s)}^2$ is proportional to the spin $s$ as $M_{(s)}^2 \propto (s-1)$ for so$(3)_R$ triplet fields.

This paper is organized as follows;
In the next section, we review the $\mathcal{N}=3$ holography proposed in \cite{Creutzig:2014ula} and introduce the marginal deformations of \cite{Hikida:2015nfa}. We also give the final result of the mass formula in \eqref{Msp} and \eqref{Msm} below.
In section \ref{lowspin} we compute the Higgs masses for low spin so$(3)_R$ singlet fields by using a brute force method. 
In section \ref{spins} we generalize the results for generic higher spin fields and also so$(3)_R$ triplet fields by making use of a free ghost system. We adopt two different ways of computation.
In section \ref{bulk} we interpret the boundary computation in terms of bulk theory. From the interpretation we deduce the correction terms at the higher order  of the deformation parameter. 
We conclude this paper and discuss future problems in section \ref{conclusion}.
In appendix \ref{BHS} we construct the free ghost realization of higher spin superalgebra and $\mathcal{N}=3$ superconformal subalgebra.
In appendix \ref{basics} we summarize the useful properties of embedding formulation for tensor fields on AdS$_{d+1}$.

\section{$\mathcal{N}=3$ holography and a summary of results}
\label{summary}

The proposal in \cite{Creutzig:2014ula,Hikida:2015nfa} includes a duality between  a 3d Prokushkin-Vasiliev theory in \cite{Prokushkin:1998bq} and a large $N$ limit of 2d coset type model. The special feature of the duality lies in the $\mathcal{N}=3$ supersymmetry.
The 3d Prokushkin-Vasiliev theory  has  extended supersymmetry only if we choose a specific mass parameter \cite{Prokushkin:1998bq,Henneaux:2012ny}. We associate U$(2M)$ Chan-Paton (CP) factor to the fields but with a U$(M)$ invariant condition, and the higher spin theory with these conditions has $\mathcal{N}=3$ supersymmetry.
The 2d coset model with $\mathcal{N}=3$ superconformal symmetry is given by the critical level model 
\begin{align}
 \frac{\text{su}(N+M)_{N+M} \oplus \text{so}(2NM)_1}{\text{su}(N)_{N+2M} \oplus
\text{su}(M)_{M+2N} \oplus \text{u}(1)_{\kappa} } \, , 
\label{coset}
\end{align}
where $\kappa = 2NM(N+M)^2$ and the central charge is
\begin{align}
c= \frac{3}{2} MN \, .
\end{align}
In order to compare the classical gravity theory, we need to take the large $N$ limit.
The proposal was confirmed by the comparison of one-loop partition function and symmetry algebra at low spins, see \cite{Beccaria:2013wqa,Creutzig:2014ula,Candu:2014yva}.

In \cite{Hikida:2015nfa} we have studied a marginal deformation of the coset model and its interpretation in dual higher spin theory. In particular, we computed the Higgs mass of a spin 2 field both from the CFT and the higher spin theory. The aim of this paper is to compute the Higgs masses of spin $s \geq 2$ fields in the higher spin gauge theory.
In this section, we summarize the necessary information on the $\mathcal{N}=3$ holography and the mass formula obtained in this paper.

\subsection{Higher spin superalgebra}

We start from the gauge algebra of the Prokushkin-Vasiliev theory with extended supersymmetry. The higher spin theory can be defined by modifying the $\mathcal{N}=2$ higher spin gauge theory with U$(M')$ CP factor in \cite{Prokushkin:1998bq}. 
Here the U$(M')$ CP factor just means that the fields take $M ' \times M'$ matrix values.
The holography with the higher spin gauge theory was proposed in \cite{Creutzig:2011fe,Gaberdiel:2013vva,Creutzig:2013tja}.
The theory includes  gauge fields with spin $s=1,3/2,2,5/2,\cdots$, which can be described by a Chern-Simons gauge theory \cite{Blencowe:1988gj}. Moreover, there are also matter fields with masses parametrized by $\lambda$. The higher spin gauge theory in \cite{Creutzig:2014ula,Hikida:2015nfa} is then obtained by a $\mathbb{Z}_2$ truncation with $\lambda =1/2$.%
\footnote{The holography with the truncation and $M' = 1$ was proposed in \cite{Beccaria:2013wqa}, where the higher spin theory has $\mathcal{N}=1$ supersymmetry. A different $\mathcal{N}=1$ holography was already conjectured in \cite{Creutzig:2012ar}.}
For the $\mathcal{N}=3$ holography, we assign U$(2M)$ CP factor and the U$(M)$ invariant condition as mentioned above.

The gauge algebra can be defined by using  $y_\alpha$ $(\alpha = 1,2)$ and $\hat k$ satisfying
\begin{align}
[y_\alpha , y_\beta ] = 2 i \epsilon_{\alpha \beta} (1 - (1 - 2 \lambda) \hat k) \, , \quad
\hat k^2 = 1 \, , \quad \{ \hat k , y_\alpha \} = 0 \, . 
\end{align}
We denote the algebra generated by these variables as $sB[\lambda]$. The gauge algebra shs$_{M'}[\lambda]$ for the $\mathcal{N}=2$ higher spin gauge theory with U$(M')$ CP factor is defined as
\begin{align}
 sB_{M'} [\lambda] \equiv sB [\lambda] \otimes \mathcal{M}_{M'} = \mathbb{C} \oplus \text{shs}_{M'} [\lambda] \, .
 \label{shs}
\end{align}
Here $\mathcal{M}_{M'}$ denotes the $M' \times M'$ matrix algebra and $\mathbb{C}$ represents the central element.
The bosonic subalgebra with $M'=1$ and in the $\hat k=1$ subsector is known as hs$[\lambda]$.
At $\lambda = 1/2$, the commutator of $y_\alpha$ does not involve $\hat k$ anymore, and the algebra can be truncated by assigning the invariance under $\hat k \to - \hat k$. The truncated algebra may be called as $\text{shs}^T_{M'}[1/2]$.

The matrix algebra $\mathcal{M}_{M'}$ with $M' = 2^n$ can be generated by the Clifford elements $\phi^I$ $(I = 1, 2, \cdots, 2n+1)$ satisfying $\{\phi^I , \phi^J \} = 2 \delta^{IJ}$. This indicates that the truncated algebra $\text{shs}^T_{M'}[1/2]$ includes osp$(2n+1|2)$ subalgebra \cite{Prokushkin:1998bq,Henneaux:2012ny,Creutzig:2014ula,Candu:2014yva}
\begin{align}
T_{\alpha \beta} = \{y_\alpha , y_\beta \} \, , \quad
Q_\alpha^I = y_\alpha \otimes \phi^I \, , \quad
M^{IJ} = [\phi^I , \phi^J]  \, . 
\end{align}
In our case, we associate U$(2M)$ CP factor and require U$(M)$ invariant condition. The subalgebra $\text{shs}^T_{2}[1/2]$ survives the invariant condition, thus the theory can be seen to have $\mathcal{N}=3$ supersymmetry.
We consider the bosonic subsector of gauge fields based on $\text{shs}^T_{2}[1/2]$. Decomposing the U$(2)$ of the CP factor as U$(1) \times$SU$(2)$, the trace part gives so$(3)_R$ singlet higher spin fields $A^{0}_{(s)}$ and the SU$(2)$ part gives so$(3)_R$ triplet higher spin fields $A^{i}_{(s)}$ with $i=1,2,3$. 
We would like to compute the masses of these fields after the deformation breaking the higher spin symmetry.
Notice that $A^{0}_{(2)}$ and $A^i_{(1)}$ should be kept massless since they are a part of generators for unbroken $\mathcal{N}=3$ supersymmetry.

\subsection{Dual CFT}

We are interested in a large $N$ limit of the symmetry algebra in the dual CFT.
Let us denote the gauge algebra of the higher spin theory as  $g_\text{hs}$.
Near the AdS boundary, the symmetry algebra
is enhanced to be a W-algebra obtained  by a Hamiltonian reduction of affine $g_\text{hs}$ algebra as explained in \cite{Henneaux:2010xg,Campoleoni:2010zq,Campoleoni:2011hg}. A claim in \cite{Gaberdiel:2011wb} is that at the large $N$ limit we can truncate the W-algebra into so called ``wedge'' subalgebra 
consistently, and the subalgebra is identical to the original higher spin algebra  $g_\text{hs}$. 
We represent the CFT currents as $J^{(s,a)} (z)$ with $a=0,1,2,3$ (or their mode expansions $J^{(s,a)}_n$ with $n \in \mathbb{Z}$), which are dual to the higher spin fields $A^{(a)}_s$ 
introduced above. 
From the above argument, the wedge subalgebra generated by $J^{(s,a)}_n$ with $|n| < s$
should be given by shs$^T_2[1/2]$ at the large $N$ limit.

It will be useful to realize the higher spin algebra shs$^T_2[1/2]$ by free ghost system as in appendix \ref{BHS}.
Since we are interested in the bosonic subsector, we only need to include one type of ghost system, say, $(b_A , c_A)$ with $A=1,2$. 
The operator product is 
\begin{align}
 b_A (z) c_B (w) \sim \frac{\delta_{AB}}{z-w} \, , 
 \label{freeghost}
\end{align}
and the conformal weights are 
\begin{align}
(h_+ , h_-) = \left(\frac{1+ \lambda}{2}  , \frac{1 - \lambda}{2}\right) \, .
\label{hpm}
\end{align}
We are interested in only the case with $\lambda = 1/2$, but
 we keep $\lambda$ generic unless necessary.
The integer spin $s$ currents are then given as \cite{Bergshoeff:1991dz}
\begin{align}
 [ V_{\lambda} ^{(s)}(z) ]_{AB} = \sum_{i=0}^{s-1} a^i (s , \lambda + 1) \partial^{s-1-i}
    \{ (\partial^i b_A) c_B\}  
\end{align} 
with
\begin{align}
 a^i (s, \lambda) = \begin{pmatrix} s-1 \\ i \end{pmatrix} 
 \frac{(-\lambda - s + 2)_{s-1-i}}{(s+i)_{s-1-i}} \quad (0 \leq i \leq s-1) \, .
\end{align}
Here we have used the following notation as  
\begin{align}
(a)_n = \frac{\Gamma (a + n)}{\Gamma(a)} = a (a + 1) \cdots (a + n -1) \, .
\label{Poch}
\end{align}

We examine two type of currents $J^{(s)}_+ (z) \equiv J^{(s,0)} (z)$ and $J^{(s)}_- (z) \equiv J^{(s,3)} (z)$ since the properties of the other so$(3)_R$ charged currents $J^{(s,a)} (z)$ with $a=1,2$ can be deduced from those with $a=3$. The properties of the wedge subalgebra for $J^{(s)}_\pm (z)$ at the large $N$ limit can be obtained from those of 
\begin{align}
V_{\lambda, \pm} ^{(s)}(z) =  [ V_{\lambda} ^{(s)}(z) ]_{11} \pm [ V_{\lambda} ^{(s)}(z) ]_{22} \, . \label{Vpm}
\end{align}
Here the normalization of the currents is%
\footnote{This information is obtained from the classical gravity theory as in \cite{Ammon:2011ua}. This is a property outside the wedge subalgebra, so the quantity cannot be computed with  the currents introduced in \eqref{Vpm}. See, for instance, \eqref{lhss} below.}
\begin{align}\label{2pt}
&\langle J^{(s)}_\pm(z_1) J^{(s)}_\pm(z_2) \rangle = - \frac{(2s-1) c N_s}{6} \frac{1}{z_{12}^{2s}} \, , 
\\
&N_s =  \frac{3  \sqrt{\pi} \Gamma(s)}{4^{s-1} (\lambda^2 - 1)\Gamma (s + \frac12)}\frac{\Gamma (s-\lambda) \Gamma (s +\lambda)}{ \Gamma(1-\lambda) \Gamma(1+\lambda) } \, ,
\label{norm}
\end{align}
where the central charge is \cite{Henneaux:2010xg,Campoleoni:2010zq}
\begin{align}
c = \frac{3}{2G_N}
\label{central}
\end{align}
with Newton's constant $G_N$.

The higher spin theory includes two complex scalar fields and two Dirac spinor fields along with higher spin gauge fields.
Due to the U$(2M)$ CP factor and the U$(M)$ invariant condition, the single particle states of the matter fields take $2 \times 2$ matrix values. These matter fields are dual to operators denoted  as $\mathcal{O}^{A \bar B}_\pm (z,\bar z)$ and $\mathcal{F}^{A \bar B}_\pm (z, \bar z)$ with $A , \bar B = 1,2$.%
\footnote{We suppress the argument of $\bar z$ in the following.
 }
The definition of these states in \cite{Hikida:2015nfa} leads to the following relation of complex conjugation as
\begin{align} 
  \mathcal{O}^{11}_\pm (z ) =  \bar{\mathcal{O}}^{22}_\pm (z ) \, , \quad 
  \mathcal{O}^{12}_\pm (z ) =  \bar{\mathcal{O}}^{21}_\pm (z ) \, , \quad 
  \mathcal{F}^{11}_\pm (z ) =  \bar{\mathcal{F}}^{22}_\pm (z ) \, , \quad 
  \mathcal{F}^{12}_\pm (z ) =  \bar{\mathcal{F}}^{21}_\pm (z ) \, .
  \label{OFcpx}
\end{align}
We choose the boundary condition of the matter fields such that  $\mathcal{O}^{A \bar B}_\pm (z , \bar z)$ have the conformal weights $(h_\pm,h_\pm)$ and $\mathcal{F}^{A \bar B}_\pm (z)$ have  $(h_\pm, h_\mp)$, where  $h_\pm$ are defined in \eqref{hpm} with $\lambda = 1/2$. For simplicity we denote the operators as
\begin{align}
  \label{OF}
	& \mathcal{O}^{1}_+ (z) \equiv  \mathcal{O}^{11}_+ (z) \, , \quad
	 \mathcal{O}^{2}_+ (z) \equiv  \mathcal{O}^{12}_+ (z) \, , \quad
	 \mathcal{O}^{1}_- (z)  \equiv  \mathcal{O}^{2 2}_- (z) \, ,   \quad
	 \mathcal{O}^{2}_- (z)  \equiv  \mathcal{O}^{2 1}_- (z) \, ,   \\
	& \mathcal{F}^{1}_+ (z)  \equiv  \mathcal{F}^{11}_+ (z) \, , \quad
	 \mathcal{F}^{2}_+ (z)  \equiv  \mathcal{F}^{12}_+ (z) \, , \quad
	 \mathcal{F}^{1}_- (z)  \equiv  \mathcal{F}^{2 2}_- (z)  \, , \quad
	 \mathcal{F}^{2}_- (z)  \equiv  \mathcal{F}^{2 1}_- (z) \, . \nonumber
\end{align}
The three point functions are computed from the bulk higher spin gauge theory as \cite{Chang:2011mz,Ammon:2011ua,Creutzig:2012xb,Moradi:2012xd}
\begin{align}
 \label{3pt}
&\langle \mathcal{O}^a_\pm (z_1) \bar{\mathcal{O}}^a_\pm (z_2) J^{(s)}_\eta (z_3) \rangle 
 = A_\pm (s , \lambda)\left( \frac{z_{12}}{z_{13}z_{23}} \right)^s 
 \langle \mathcal{O}^a_\pm (z_1) \bar{\mathcal{O}}^a_\pm (z_2)  \rangle \, , \\
&\langle \mathcal{F}^a_\pm (z_1) \bar{\mathcal{F}}^a_\pm (z_2) J^{(s)}_\eta (z_3) \rangle 
 = A_\pm (s , \lambda) \left( \frac{z_{12}}{z_{13}z_{23}} \right)^s 
 \langle \mathcal{F}^a_\pm (z_1) \bar{\mathcal{F}}^a_\pm (z_2)  \rangle 
  \nonumber
\end{align}
with%
\footnote{The holography requires that either of $A_+$ or $A_-$ has the factor $(-1)^s$, and here we choose to put the factor in $A_-$, see, e.g., \cite{Ammon:2011ua}. Moreover, the so$(3)_R$ charges can be read from the action of the Pauli matrix $\sigma^a$. In our choice, $\mathcal{O}^a_+$ and $\bar{\mathcal{O}}^a_-$ has $+1$ eigenvalue of $\sigma^3$ and  $\bar{\mathcal{O}}^a_+$ and $\mathcal{O}^a_-$ has $-1$ eigenvalue of $\sigma^3$. The  so$(3)_R$  charge assignment is similar for the fermionic operators.}
\begin{align} 
A_+ (s , \lambda) = \frac{\Gamma(s)^2}{\Gamma(2s-1)} \frac{\Gamma(s + \lambda)}{\Gamma (1 + \lambda)} \, , \quad
A_- (s , \lambda) = \eta (-1)^s \frac{\Gamma(s)^2}{\Gamma(2s-1)} \frac{\Gamma(s - \lambda)}{\Gamma (1 - \lambda)} \, .
\label{Npm}
\end{align}
These correlation functions can be  reproduced by using a free ghost system as shown in \cite{Moradi:2012xd}.
 In principal, it is possible to compute them directly using the coset model \eqref{coset} with finite $N$ and then taking the large $N$ limit. However, the computation would be quite complicated, and it is convenient to use the classical gravity theory and the free ghost system at the limit.

\subsection{Marginal deformation and Higgs masses}

In order to compare to superstring theory with finite string tension, we need to break the higher spin symmetry. As in \cite{Hikida:2015nfa} we deform the bulk higher spin theory by changing boundary conditions of the U$(M)$ singlet matter fields with keeping $\mathcal{N}=3$ supersymmetry. 
The change of boundary conditions is dual to the following double-trace type deformation of the dual CFT as \cite{Hikida:2015nfa}
\begin{align}
\label{def}
 &\Delta S = - f \int d^2 w \mathcal{T} (w , \bar w) \, , \\
& \mathcal{T} = \sum_{a=1}^2  \frac{(-1)^{a-1}}{2} \left[ \mathcal{O}^a_+ \mathcal{O}^a_- +\bar{\mathcal{O}}^a_-  \bar{\mathcal{O}}^a_+ + \mathcal{F}^a_+ \mathcal{F}^a_- +  \bar{\mathcal{F}}^a_- \bar{\mathcal{F}}^a_+ \right] \, .
\label{defop}
\end{align}
The deformation is expected to break higher spin symmetry generically, and the corresponding currents are not conserved any more as
\begin{align}
 \bar \partial J_\pm^{(s)}(z) = \mathcal{K}_\pm^{(s-1)} (z)
 \label{delj}
\end{align}
with $\mathcal{K}_\pm^{(s-1)} (z)$ as spin $(s-1)$ operators.

The aim of this paper is to compute the masses of spin $s$ fields, which are generated due to the symmetry breaking. A direct method is to compute the one-loop corrections of spin $s$ propagators as was done in \cite{Porrati:2001db,Duff:2004wh,Aharony:2006hz} for spin 2 fields. However the computation would be quite complicated. Instead of the direct way, we compute the masses from the viewpoint of dual CFT. In other words, we compute the masses by making use of the AdS isometry, which is the same as the conformal symmetry of the boundary CFT.
The map is known between the mass of bulk spin $s$ field and the conformal dimension $\Delta^{(s)}_\pm$  of dual spin $s$ current $J_\pm^{(s)} (z)$ as  
\begin{align}
 M^2_{(s,\pm )} = \Delta^{(s)}_\pm (\Delta^{(s)}_\pm - 2 ) - s (s-2) \, .
 \label{dictionary}
\end{align}
This formula reduces to 
\begin{align}
 M^2_{(s,\pm)} =  2 (s-1 )(\Delta^{(s)}_\pm - s)
 \label{dictionary2}
\end{align}
at the first order of  the anomalous dimension $\Delta^{(s)}_\pm - s$.
Using the map
we can compute the mass of spin $s$ field from the dual CFT.

In the rest of the sections, we compute the anomalous dimensions using various methods. Before going into the details of the computation, we summarize our results on the mass formula here.
Our results are at the first order of $1/c$ since we heavily use the classical bulk gravity theory and the free ghost system \eqref{freeghost}. However, we can obtain the Higgs masses at the full order of the deformation parameter $f$ in \eqref{def}. 
As we saw above, there are so$(3)_R$ singlet and triplet higher spin fields.
The so$(3)_R$ singlet  fields do not receive any corrections as
\begin{align}
 M^2_{(s,+)} = 0  \quad (s=2,3,4,\ldots)\, .
 \label{Msp}
\end{align}
The result may be expected because the deformation operator \eqref{defop} is of the same type as the one in \cite{Gaberdiel:2013jpa}, where the authors considered the deformation preserving the higher spin symmetry at the leading order of $1/c$ and the deformation parameter $f$.
For the so$(3)_R$ triplet fields, the mass formula is obtained as
\begin{align}
  M_{(s,-)}^2  = \frac{f^2}{(1 + f^2)^2}  \frac{12  (s-1)  }{c} \quad (s=1,2,3,4,\ldots)\, ,
   \label{Msm}
 \end{align}
where the central charge is related to the Newton constant $G_N$ as \eqref{central}.
It reproduces the result in \cite{Hikida:2015nfa} for $s=2$.
More detailed examination of these results will be given in the concluding section.

\section{The examples of low spin currents}
\label{lowspin}

We start from the simple examples with $s=2,3,\ldots$ and of the so$(3)_R$ singlet, which are given by $J_+^{(s)}(z)$.
These higher spin currents generate the hs$[\lambda]$ bosonic subalgebra.
We would like to deform the theory as in \eqref{defop}, but here we consider a simpler version as
\begin{align}
\mathcal{T}_\lambda = \frac12  \left [\mathcal{O}_+ \mathcal{O}_ - +  \bar{\mathcal{O}}_-  \bar{\mathcal{O}}_+ \right] \, .
\label{defop2}
\end{align}
Denoting the corresponding state as $| \mathcal{T}_\lambda \rangle$,
the eigenvalues of current zero modes are written as
\begin{align}
J_{+,0}^{(s)} |\mathcal{T}_\lambda \rangle = (A_+(s,\lambda) + A_-(s , \lambda)) | \mathcal{T}_\lambda \rangle \, .
\label{hscharges}
\end{align}
Here $A_+(s,\lambda)$ is defined in \eqref{Npm}.
For $\lambda = 1/2$, we can identify $\mathcal{O}_\pm$ as $\mathcal{O}^a_\pm$ $(a=1,2)$ in \eqref{OF}. We can use the same results for the fermionic operators  $\mathcal{F}^a_\pm$ $(a=1,2)$ in \eqref{OF}. 

After the deformation, the divergence of currents can be written as
(see, e.g., \cite{Gaberdiel:2013jpa,Hikida:2015nfa})
\begin{align}
\bar \partial J_+^{(s)}(z) = 2 \pi f \sum_{l=0}^{s-1} \frac{(-1)^l}{l!}
 (L_{-1})^l J^{(s)}_{+,-s+l+1} \mathcal{T}_\lambda (z) 
 \label{delj2}
\end{align}
with $L_{n} = J^{(2)}_{+,n}$. The right hand side vanishes for $s=2$ since the eigenvalue of $J^{(2)}_{+,0}$ is one in our case. This implies that the mass is not generated for the usual graviton field as in \eqref{Msp}.
The square of the left hand side of \eqref{delj2} can be computed as
\begin{align}
 |\bar \partial J_+^{(s)}|^2 \equiv \langle 0 | J^{(s)}_{+,s} \bar L_{1} \bar L_{-1} J^{(s)}_{+,-s} | 0 \rangle
  =  - (\Delta^{(s)}_+ - s)  \frac{(2s-1) N_s c}{6} \, ,
  \label{lhss}
\end{align}
where \eqref{2pt} is used. Therefore, if we can compute the square  of the right hand side of \eqref{delj2}, then we can read off the anomalous dimension for $J_+^{(s)}(z) $ from this expression.
In this section, we compute the square of the right hand side explicitly for low spin examples with $s=3,4$.

\subsection{Spin 3 current}
\label{spin3}

For the spin 3 current $W(z) \equiv J^{(3)}_+ (z)$, the divergence of current in \eqref{delj2} becomes
\begin{align}
 \bar \partial W = 2  \pi f \left(W_{-2} - L_{-1} W_{-1} + \frac12 L_{-1}^2 W_{0} \right) \mathcal{T}_\lambda \, .
 \label{noncons}
\end{align}
The problem is now to compute the square of the right hand side explicitly.
For the purpose we need  the commutation relations among the mode expansions of higher spin currents, which are given as%
\footnote{We replaced $N_3$ in the commutation relations of the paper \cite{Gaberdiel:2012ku} by $- N_3$, effectively this amounts to a rescaling of the spin four field compared to \cite{Gaberdiel:2012ku}. After this replacement, the current-current two point function becomes \eqref{2pt} with positive coefficient for $0 < \lambda < 1$. }
 \begin{align} 
&  [L_m , L_n] = (m-n) L_{n+m} + \frac{c}{12} m (m^2 -1) \delta_{m+n} \, , \quad
 [L_m , W_n] = (2 m - n) W_{m+n} \, , \nonumber \\
&  [W_m , W_n] = 2 (m-n) U_{m+n} - \frac{N_3}{12} (m - n) (2 m^2 + 2 n^2 - mn - 8) L_{m+n}  \label{cm} \\ & \qquad \qquad  \quad - \frac{8 N_3} {(c + \frac{22}{5})} (m-n) \Lambda^{(4)}_{m+n} - \frac{N_3 c}{144} m (m^2 -1 ) (m^2 - 4) \delta_{m+n} 
\nonumber
\end{align}
with $U$ as the spin 4 field $J^{(4)}_{+}$ and $\Lambda^{(4)}_{m}$ as a composite operator made with $L_n$. 
The terms proportional to $\Lambda^{(4)}_{m}$ are subleading for large $c$ and hence are neglected, and the expression of $N_3$ is given in \eqref{norm}. 

We compute the anomalous dimension by comparing the norm of the both side of \eqref{noncons} as%
\footnote{The square $|\bar \partial W|^2$ is defined as in \eqref{lhss}. The same notation will be used below as well.}
\begin{align}
  \label{square}
 |\bar \partial W|^2 = (2 \pi f)^2 & \langle \mathcal{T}_\lambda | \left(W_2 - W_1 L_1 + \frac12 W_0 (L_1)^2 \right)  \\ & \qquad \qquad   \cdot
 \left (W_{-2} - L_{-1} W_{-1} + \frac12 (L_{-1}) W_0 \right) | \mathcal{T}_\lambda \rangle \, .   \nonumber
\end{align}
The following eigenvalues are introduced as
\begin{align}
 L_0 | \mathcal{T}_\lambda \rangle  = h  | \mathcal{T}_\lambda \rangle  \, , \quad
 W_0  | \mathcal{T}_\lambda \rangle  = w | \mathcal{T}_\lambda \rangle  \, , \quad
 U_0  | \mathcal{T}_\lambda \rangle  = u | \mathcal{T}_\lambda \rangle  \, .
\end{align}
Then we have
\begin{align}
\frac14 & \langle \mathcal{T}_\lambda | W_0 (L_1)^2 (L_{-1})^2  W_0 | \mathcal{T}_\lambda \rangle  = (1 + 2 h) h w^2 \langle \mathcal{T}_\lambda | \mathcal{T}_\lambda \rangle \, , \nonumber \\
 - \frac12  &\langle \mathcal{T}_\lambda | W_0 (L_1)^2 L_{-1} W_{-1} | \mathcal{T} _\lambda\rangle
 =   - \frac12 \langle \mathcal{T} _\lambda| W_1 L_1 (L_{-1})^2  W_0 | \mathcal{T}_\lambda \rangle
  = - 3 w^2 (1 + 2 h)\langle \mathcal{T}_\lambda | \mathcal{T} _\lambda\rangle \, , \nonumber  \\
 \frac12 & \langle \mathcal{T}_\lambda | W_2 (L_{-1})^2 W_0 | \mathcal{T}_\lambda \rangle
 =  \frac12\langle \mathcal{T}_\lambda | W_0 (L_{1})^2 W_{-2}  | \mathcal{T}_\lambda \rangle
  = 6 w^2 \langle \mathcal{T}_\lambda | \mathcal{T}_\lambda \rangle \, , \\
& \langle \mathcal{T}_\lambda | W_1 L_1 L_{-1} W_{-1} | \mathcal{T} _\lambda\rangle
  = \left( 2 (h+1)  \left(4 u + \frac{N_3}{2} h \right) + 9 w^2 \right)  \langle \mathcal{T} _\lambda | \mathcal{T} _\lambda\rangle \, , \nonumber \\  
 & \langle \mathcal{T} _\lambda| W_2 W_{-2}| \mathcal{T}_\lambda \rangle 
  = 4 (2 u - N_3 h) \langle \mathcal{T}_\lambda | \mathcal{T}_\lambda \rangle \, , \nonumber  \\
   - & \langle \mathcal{T}_\lambda | W_2 L_{-1} W_{-1}| \mathcal{T}_\lambda \rangle 
   = - \langle \mathcal{T}_\lambda | W_1 L_1 W_{-2}| \mathcal{T}_\lambda \rangle 
  = - 4 \left(4 u + \frac{N_3}{2} h \right) \langle \mathcal{T} _\lambda| \mathcal{T}_\lambda \rangle \, .  \nonumber
  \end{align}
  Thus the right hand side of \eqref{square} leads to
  \begin{align} 
   |\bar \partial W|^2 =  
   [u (8h - 16) + N_3 h (h-7) + w^2 (2 h^2 - 11 h + 15) ]  (2 \pi f)^2  \langle \mathcal{T } _\lambda| \mathcal{T}_\lambda \rangle 
   = 0 \, .
\label{rhs}
  \end{align}
Here we have used
\begin{align}
 h = 1 \, , \quad w = \lambda \, , \quad  u = \frac{3}{5} (1 + \lambda^2) \, , \quad N_3 =  \frac{1}{5} (\lambda^2 - 4) \, ,
\end{align}
which come from \eqref{hscharges}.
Therefore  at the leading order of $1/c$ and $f^2$ we find as in \eqref{Msp}
\begin{align}
 M^2_{(3)} = 4 (\Delta - 3) = 0 \, ,
\end{align}
where we have used \eqref{dictionary2} and \eqref{lhss}.

 \subsection{Spin $4$ current}
\label{spin4}

If we want to compute the deformation for the spin 4 current we need to know its commutation relations again. The operator product of the field $U(z)$ with itself has only poles of even order and the coefficients can only be normally ordered polynomials in $T(z), \partial T(z), \partial^2 T(z), \partial^3 T(z), \partial^4 T(z),  W(z), U(z), \partial U(z), \partial^2 U(z), Y(z)$ where $Y(z)$ is the spin 6 current $J_+^{(6)}$. As explained in \cite{Blumenhagen:1990jv} there are various relations between structure constants. Especially the one of a normally ordered polynomials of type $:\partial^j T(z) X(z):$ for some primary field $X(z)$ is related to the structure constant for $X(z)$. In  \cite{Blumenhagen:1990jv} an explicit formula for this relation is given, it is very lengthy, but one can easily extract that they behave as $1/c$ for large $c$ and are thus negligible for our considerations. 
In \cite{Gaberdiel:2012yb} terms of the commutator of the spin four field modes with themselves were computed, the term corresponding to $:W(z)W(z):$ is also subleading for large $c$ and hence their results imply that
\begin{align}\nonumber
[U_m, U_n] =\ & 3(m-n)Y_{m+n}-f_U(m, n) U_{m+n}+f_L(m, n) L_{m+n} +\\  \nonumber
&- \frac{cN_4}{4320} m(m^2-1)(m^2-4)(m^2-9) \delta_{m+n, 0}+\\
&+P(m, n, T, T', T'', T''', T'''', W, U, U', U'') \, , \\  \nonumber
f_U(m, n) =\ & n_{44}(m-n)(m^2-mn+n^2-7) \, , \\  \nonumber
f_L(m, n) =\ & -\frac{N_4}{360}(m-n)\left(108-39m^2+3m^4+20mn-2m^3n-39n^2+\right. \\
&\qquad\qquad\left. +4m^2n^2-2mn^3+3n^4 \right) \, , \nonumber
\end{align}
where $P$ denotes modes of a normally ordered polynomial in the indicated fields. It is subleading and can be neglected for our computations. 
For general $\lambda$, the involved constants can be read off by comparing to \cite{Gaberdiel:2012yb}. They are
\begin{align}
n_{44}=\frac{(\lambda^2-19)}{30} \, , \quad N_4=-\frac{3}{70}(\lambda^2-4)(\lambda^2-9)  
\end{align}
and the charges of the zero modes $Y_0$ and $U_0$ on $|\mathcal T_\lambda \rangle$ are obtained from \eqref{hscharges}. They are
\begin{align}
y=\frac{5}{42}(8+15\lambda^2+\lambda^4) \, , \quad u=\frac{3}{5}(1+\lambda^2) \, .
\end{align}
We need to compute to leading order in $1/c$ of 
\begin{equation}
\begin{split}
\text{rhs}:=\ & \langle \mathcal{T}_\lambda| \left(U_3-U_2L_1+\frac{1}{2}U_1\left(L_1\right)^2-\frac{1}{6}U_0\left(L_1\right)^3\right)\\
&\quad\quad
\cdot  \left(U_{-3}-L_{-1}U_{-2}+\frac{1}{2}\left(L_{-1}\right)^2U_{-1}-\frac{1}{6}\left(L_{-1}\right)^3U_{0}  \right)|\mathcal{T}_\lambda\rangle \, .
\end{split}
\end{equation}
This computation is straightforward and lengthy. 
First define 
\begin{align}
Z_n:= \langle \mathcal{T}_\lambda| U_n U_{-n}|\mathcal{T}_\lambda\rangle = \begin{cases} \left( 6ny-f_U(n, -n)u+f_L(n, -n)h\right) \langle \mathcal{T}_\lambda|\mathcal{T}_\lambda\rangle \quad &(n\neq 0) \, , \\
u^2 \langle \mathcal{T}_\lambda|\mathcal{T}_\lambda\rangle  \quad &(n=0) \, . \end{cases}
\end{align}
We get
\begin{align}
 \langle \mathcal{T}_\lambda| U_3 U_{-3}|\mathcal{T}_\lambda\rangle &= Z_3  \, ,\nonumber \\
\langle \mathcal{T}_\lambda| U_3 L_{-1}U_{-2}|\mathcal{T}_\lambda\rangle &= 6Z_2 \, ,\nonumber\\
\langle \mathcal{T}_\lambda| U_3 \left(L_{-1}\right)^2U_{-1}|\mathcal{T}_\lambda\rangle &= 30Z_1 \, ,\nonumber\\
\langle \mathcal{T}_\lambda| U_3 \left(L_{-1}\right)^3U_{0}|\mathcal{T}_\lambda\rangle &= 120 Z_0 \, ,\nonumber\\
\langle \mathcal{T}_\lambda| U_2L_1 L_{-1}U_{-2}|\mathcal{T}_\lambda\rangle &= 25Z_1+2(h+2)Z_2 \, ,\\
\langle \mathcal{T}_\lambda| U_2L_1 \left(L_{-1}\right)^2U_{-1}|\mathcal{T}_\lambda\rangle &= 10(2h+3)Z_1+80Z_0 \, ,\nonumber\\
\langle \mathcal{T}_\lambda| U_2L_1\left(L_{-1}\right)^3U_{0}|\mathcal{T}_\lambda\rangle &=  120 (h+1)Z_0 \, ,\nonumber\\
\langle \mathcal{T}_\lambda| U_1\left(L_1\right)^2 \left(L_{-1}\right)^2U_{-1}|\mathcal{T}_\lambda\rangle &= 4(2h+3)(h+1)Z_1+128(h+1)Z_0 \, ,\nonumber\\
\langle \mathcal{T}_\lambda| U_1\left(L_1\right)^2 \left(L_{-1}\right)^3U_{0}|\mathcal{T}_\lambda\rangle &= 48(2h^2+3h+1)Z_0 \, ,\nonumber\\
\langle \mathcal{T}_\lambda| U_0\left(L_1\right)^3 \left(L_{-1}\right)^3U_{0}|\mathcal{T}_\lambda\rangle &= 24h(2h^2+3h+1)Z_0 \, .\nonumber
\end{align}
So that with $h=1$
\begin{align}
\text{rhs} = Z_3-6Z_2+15Z_1-20Z_0\, .
\end{align}
Plugging $n_{44}$ and $N_4$ into the expresions for $f_L(n, m)$ and $f_U(n, m)$, we get
\begin{align}
\text{rhs}= 36y+4(\lambda^2-19)u+\frac{18}{35}(\lambda^2-4)(\lambda^2-19)-20u^2 
\end{align}
and it turns out that independent of $\lambda$ this expression vanishes identically
\begin{equation}\label{rhs4}
\text{rhs}=0 \, .
\end{equation}
Therefore at the leading order of $1/c$ and $f^2$ we again find as in \eqref{Msp}
\begin{align}
 M^2_{(4)} =  6 (\Delta_+^{(4)} - 4) = 0 
\end{align}
with the use of \eqref{dictionary2} and \eqref{lhss}. 

\section{Generic spin $s$ currents}
\label{spins}

As above, we can compute the anomalous dimension of $J^{(s)}_+ $ by comparing the two ways to express $|\bar \partial J^{(s)}_+ |^2$. One way can be found in  \eqref{lhss}.
 The other way is to compute $|\mathcal{K}^{(s-1)}|^2$ using \eqref{delj}.
 The computation with the expression in \eqref{delj2} becomes complicated rapidly when we increase $s$ as seen in the examples with $s=3,4$.
 Fortunately, the right hand side in \eqref{delj2} involves only the wedge subalgebra with $J^{(s)}_{+,n}$ $(|n| < s)$, so we can utilize the free ghost system  \eqref{freeghost}.
 There are at least two merits to use the free ghost system. One is that the other expression of $|\bar \partial J^{(s)}_+|^2$ can be dealt with  more easily even for generic $s$. Another is that the analysis can be generalized to the so$(3)_R$ triplet currents simply by inserting a phase factor $-1$ as in \eqref{Vpm}. In this section, we obtain the mass formula \eqref{Msp} and \eqref{Msm} at the first order of $1/c$ and $f^2$ in two ways. The analysis on the higher order of $f^2$ is postponed to later sections.

\subsection{Direct computation}
\label{direct}

As explained above, we can utilize the generators $V_{\lambda,\pm}^{(s)} (z)$ defined in \eqref{Vpm}.
The generators have the following operator products like
\begin{align}
 &V_{\lambda,\pm}^{(s)} (z) b_1 (w) \sim \sum_{i=0}^{s-1} a^i (s , \lambda + 1) \partial^{s-1-i}_z
 \left( \frac{1}{z-w} \right) \partial^i b_1 (w) \, ,
 \label{freeOPE2}\\
 &V_{\lambda,\pm}^{(s)} (z) c_2 (w) \sim \pm (-1)^s \sum_{i=0}^{s-1} a^i (s , 1 - \lambda) \partial^{s-1-i}_z
 \left( \frac{1}{z-w} \right) \partial^i c_2 (w) \, .
 \label{freeOPE3}
 \end{align}
We can reproduce  \eqref{3pt} and \eqref{Npm} if we identify 
 $V_{\lambda,\pm}^{(s)}$ as $J^{(s)}_\pm$ and $b_1,c_2$ as $\mathcal{O}_+, \mathcal{O}_-$, see \cite{Moradi:2012xd}.  
Therefore, we have
\begin{align}
&\bar \partial V_{\lambda,\pm}^{ (s) } (z) (b_1 c_2 ) (w) \\ \nonumber
 & \quad = 2 \pi \delta^{(2)} (z-w)  \sum_{i=0}^{s-1}  [ a^i (s , \lambda + 1) 
   \partial^{s-1-i} ( \partial^i b_1 c_2 ) (w)  \pm (-1)^s
   a^i (s , 1 - \lambda ) 
    \partial^{s-1-i} ( b_1 \partial^i c_2  ) (w) ] \\
  & \quad= 2 \pi \delta^{(2)} (z-w)  \sum_{i=0}^{s-1} (1 \mp 1 )   \tilde a^i (s , \lambda + 1) 
   ( \partial^i b_1 \partial^{s-1-i} c_2 ) (w) \, .
     \nonumber
\end{align}
In the above expression, we have neglected total derivatives, which vanish after the integration over the position $w$ of deformation operator.
Here $\tilde a^i (s , \lambda )$ is given as
\begin{align}
 &\tilde a^i (s , \lambda) = \begin{pmatrix} s-1 \\ i \end{pmatrix} \frac{(-1)^i}{(s)_{s-1}}
  (\lambda - s)_i (2 - \lambda - s)_{s - 1 - i} \, , \label{tildea0}
\end{align} 
  and the  identities (see \cite{Bergshoeff:1991dz})
\begin{align}
& \sum_{i=0}^{s-1} a^i (s , \lambda) \partial^{s - 1 - i} ((\partial^i  A) B) 
 = \sum_{i=0}^{s-1} \tilde a (s , \lambda) (\partial^i A) (\partial^{s-1-i} B) \, , \\
& \tilde a ^i (s , \lambda) = (-1)^{s-1} \tilde a^{s -1 - i} (s , 2- \lambda) 
\end{align} 
are used.

With these preparations, we can compute the divergence of current  as
\begin{align}
\bar \partial J^{(s)} _\pm (z) = \pi f \sum_{i=0}^{s-1}   [   (1 \mp 1 )    \tilde a^i (s , \lambda + 1)
   ]( \partial^{s-1-i} \mathcal{O}_+ \partial^i \mathcal{O}_- ) (z)
\end{align}
after the deformation with $\frac12 \mathcal{O}_+\mathcal{O}_- (w)$.
Let us define $| \mathcal{O}_\pm\rangle$ as the eigenstate of $L_0$ with
\begin{align}
 L_0 | \mathcal{O}_\pm \rangle = h_\pm | \mathcal{O}_\pm  \rangle
\end{align} 
and $\langle \bar{\mathcal{O}}_\pm |$ as its conjugate state. Then we can show that 
\begin{align}
 \langle \bar{\mathcal{O}}_\pm | (L_1)^s (L_{-1})^s | \mathcal{O}_\pm \rangle =F(s,h_\pm)
\langle \bar{\mathcal{O}}_\pm | \mathcal{O}_\pm \rangle \, , \quad F(s,h_\pm) = \Gamma (s+1) \frac{\Gamma (2h_\pm +s)}{\Gamma (2h_\pm )}
\label{2ptss}
\end{align}
by repeatedly using the commutation relations among $L_n$ $(n=0,\pm1)$.
With the formula, we have
\begin{align}
\nonumber
 \langle  \partial^{s-1-i} \bar{\mathcal{O}}_+ \partial^i \bar{\mathcal{O}}_-  |
  \partial^{s-1-i} \mathcal{O}_+ \partial^i \mathcal{O}_-  \rangle
 =  F(s-1-i,h_+) F(i,h_-) C_+ C_-
\end{align}
with $C_\pm = \langle \bar{ \mathcal{O} }_\pm | \mathcal{O}_\pm \rangle$.
Adding  the conjugate deformation operator $\frac12 \bar{\mathcal{O}}_- \bar{\mathcal{O}}_+ (w)$, we find
\begin{align}
\nonumber
& |\bar \partial J^{(s)}_\pm  |^2 
 = (2 \pi f)^2  \sum_{i=0}^{s-1} \left\{  [ (1 \mp 1)   \tilde a^i (s , \lambda + 1)
    ]^2  F(s-1-i,h_+) F(i,h_-) \right\} \langle \mathcal{T}_\lambda | \mathcal{T}_\lambda \rangle \\
   & \quad = (2 \pi f)^2  \left[- \frac{1 \mp 1}{2} \frac{2^{4-2s} \pi^{3/2} \text{Csc} \, (\lambda \pi) \Gamma(s)}{\lambda \Gamma(1 - \lambda -s) \Gamma(1 + \lambda - s) \Gamma(-1/2 + s)} \right]  \langle \mathcal{T}_\lambda | \mathcal{T}_\lambda \rangle \, .
   \label{rhss}
\end{align} 
With \eqref{dictionary2} and \eqref{lhss} we obtain
\begin{align}
 M_{(s,\pm)}^2 &=  \frac{12 (s-1)}{(2s-1)N_s c} (2 \pi f)^2  \left[\frac{1 \mp 1}{2} \frac{2^{4-2s} \pi^{3/2} \text{Csc} \, (\lambda \pi) \Gamma(s)}{\lambda \Gamma(1 - \lambda -s) \Gamma(1 + \lambda - s) \Gamma(-1/2 + s)} \right] \langle \mathcal T _\lambda | \mathcal T _\lambda \rangle \nonumber \\
 &=   \frac{1 \mp 1}{2}  \frac{8 (1 - \lambda^2) }{c} (s-1) (2 \pi f)^2 \langle \mathcal T _\lambda | \mathcal T _\lambda \rangle \, .
\label{masss}
\end{align}
This reproduces our findings  for the so$(3)_R$ singlet fields with $s=3,4$.

For the original problem we have to set $\lambda=1/2$ and multiply the factor 4.
Furthermore, the standard kinetic term for the dual matter fields fixes the normalization as
$C_+ = C_- = 1/(2 \pi)$ \cite{Hikida:2015nfa}.
Thus the mass for the spin $s$ field dual to $J^{(s)}_\pm$ can be computed as
\begin{align}
 M_{(s,\pm)}^2  = \frac{1 \mp 1}{2}  \frac{12  (s-1)  }{c}(2 \pi f)^2
\end{align}
at the leading order of $1/c$ and $f^2$. This is the term at  the order of $f^2$ in the mass formula of \eqref{Msp} and \eqref{Msm}.

\subsection{Alternative computation}
\label{another}

In this subsection, we reproduce \eqref{rhss} in a different way of computation for the following two purposes. One is to check the computation obtained above.
Another is for a preparation of later analysis.
We will see that this way of computation has the dual gravity interpretation in terms of Witten diagram. Relying on the interpretation, we will include the corrections at the higher order of $f^2$ to the mass formula as in \eqref{Msp} and \eqref{Msm}.

Here we utilize the standard method of conformal perturbation theory as
\begin{align}
\nonumber 
& \langle \bar \partial J^{(s)}_\pm (z) \bar \partial J^{(s)}_\pm (w) e^{f \int d^2 x \mathcal{T}_\lambda} \rangle
 =  \langle \bar \partial  J^{(s)}_\pm (z) \bar \partial J^{(s)}_\pm (w) \rangle 
  + f \int d^2 x  \langle \bar \partial  J^{(s)}_\pm (z) \bar \partial J^{(s)}_\pm (w) \mathcal{T}_\lambda (x) \rangle \\
 & \quad \quad + \frac{f^2}{2} \int d^2 x \int d^2 y \langle \bar \partial  J^{(s)}_\pm (z) \bar \partial  J^{(s)}_\pm (w) \mathcal{T}_\lambda (x) \mathcal{T}_\lambda (y)  \rangle + \cdots \, .
 \label{expansion}
\end{align}
The correlation functions in the above expression are evaluated by using the non-perturbed theory. We can see that the first two terms in the right hand side vanishes.
In the third term, there are two types of contribution as
\begin{align}
I_1(x,w) =  \int d^2 x \int d^2 y \langle \bar \partial J^{(s)}_\pm (z) \mathcal{O}_+ (x) \bar{\mathcal{O}}_+(y) \rangle \langle \bar  \partial J^{(s)}_\pm (w) \mathcal{O}_- (x) \bar{\mathcal{O}}_- (y) \rangle \, , 
 \label{type1}
\end{align}
and
\begin{align}
I_2(x,w) =   \int d^2 x \int d^2 y \langle \bar \partial J^{(s)}_\pm (z) \bar \partial J^{(s)}_\pm (w) \mathcal{O}_+ (x) \bar{\mathcal{O}}_+(y) \rangle \langle  \mathcal{O}_- (x) \bar{\mathcal{O}}_- (y) \rangle \, .
 \label{type2}
\end{align}

We first consider the contribution of the type in \eqref{type1}. Taking derivative of \eqref{3pt} with respect to $\bar z_3$, we have
\begin{align}
\frac{\partial}{\partial \bar z_3} \left( \frac{1}{z_{13}z_{23}} \right)^s 
 = - \frac{2 \pi}{(s-1)!}  \left[ ( \partial^{s-1}_{z_3} \delta^{(2)} (z_{13}) ) \frac{1}{(z_{23})^{s}} 
  + ( \partial^{s-1}_{z_3} \delta^{(2)} (z_{23}) )\frac{1}{(z_{13})^{s}} \right] \, .
\end{align}
Since there is no contribution from the terms proportional to $\delta^{(2)} (z - x) \delta^{(2)} (w - x)$ for $z \neq w$, we obtain
\begin{align}
& I_1 (z,w) =  \pm 2 A_+ (s , \lambda) A_- (s , \lambda) C_+ C_- \frac{(2 \pi)^2}{[(s-1)!]^2} \\
&  \times \int d^2 x \int d^2 y  \delta^{(2)} (x - z ) \delta^{(2)} (y-w) \partial^{s-1}_{x} \partial^{s-1}_y
 \left[ \frac{1}{(y-z)^s}  \frac{1}{(x-w)^s} \frac{1}{(x-y)^{2-2s}}  \right] \frac{1}{(\bar x - \bar y)^2} \, , \nonumber
\end{align}
where we have used the invariance under the exchange of $x$ and $y$.
Evaluating the action of derivatives,
we find%
\footnote{
The second equality is checked  for $s = 2 ,3 , \ldots , 300$. \label{minus}}
\begin{align}
 & I_1 (z,w) = \pm 2  A_+ (s , \lambda) A_- (s , \lambda) C_+ C_- \frac{(2 \pi)^2}{[(s-1)!]^2} \frac{1}{(z-w)^{2s} (\bar z - \bar w)^2} \\
& \quad \times\sum_{k,\ell=0}^{s-1} \begin{pmatrix} s- 1 \\ k \end{pmatrix}
  \begin{pmatrix} s- 1 \\ \ell \end{pmatrix} (s)_{s-1-k} (2-2s)_k (2-2s+k)_\ell
   (s)_{s-1-\ell} 
   \nonumber \\
    &= \pm (2\pi)^2  
     \frac{2^{3-2s} \pi^{3/2} \text{Csc} \, (\lambda \pi) \Gamma(s)}{ \lambda \Gamma(1 - \lambda -s) \Gamma(1 + \lambda - s) \Gamma(-1/2 + s)} 
      C_+ C_-  \frac{1}{(z-w)^{2s} (\bar z - \bar w)^2} \, . \nonumber
 \end{align}
 For the total contribution, we need to multiply a pre-factor.
 First we have $f^2/2$ in \eqref{expansion}. Moreover, the exchange of $x,y$ yields a factor 2.
 From $\mathcal{T}(x ) \mathcal{T}(y ) $, we have two terms as
 \begin{align}
  \mathcal{T}(x ) \mathcal{T}(y ) =
  \frac14 \mathcal{O}_+ (x) \mathcal{O}_- (x) \bar{\mathcal{O}}_- (y) \bar {\mathcal{O}}_+ (y) +   \frac14  \bar{\mathcal{O}}_- (x) \bar {\mathcal{O}}_+ (x) \mathcal{O}_+ (y) \mathcal{O}_- (y)  + \cdots ~.
 \end{align}
Totally, we have
\begin{align}
 \frac{ f^2}{2} \cdot 2  \cdot \frac14 \cdot 2  =  \frac{ f^2}{2}  \, . 
 \label{prefactor}
\end{align}

 For the contribution of the type in \eqref{type2}, we need
 \begin{align}
\nonumber
&\bar \partial J^{ (s) }_\pm (z) \mathcal{O}_+ (x) =2 \pi \sum_{i=0}^{s-1}  a^i (s , \lambda + 1 )  \partial_z^{s-1-i} \delta^{(2)} (z-x)    \partial^i \mathcal{O}_+ (x)  \, ,
\end{align}
which comes from the free ghost computation in \eqref{freeOPE2}.
With this expression, we have
\begin{align}
& 2 (2 \pi)^2 \sum_{i=0}^{s-1}\tilde a^i(s, \lambda + 1 )^2 \langle \partial^i \mathcal{O}_+ (z)   
 \partial^i \bar{\mathcal{O}}_+ (w)  \rangle  \langle \partial^{s-1-i} \mathcal{O}_- (z)   
 \partial^{s-1-i} \bar{\mathcal{O}}_- (w)  \rangle \\
& \qquad  \qquad  = -  (2 \pi)^2
     \frac{2^{3-2s} \pi^{3/2} \text{Csc} \, (\lambda \pi) \Gamma(s)}{\lambda \Gamma(1 - \lambda -s) \Gamma(1 + \lambda - s) \Gamma(-1/2 + s)} 
C_+ C_-   \frac{1}{(z-w)^{2s} (\bar z - \bar w)^2} \, , \nonumber
\end{align}
where we have used \eqref{2ptss}.
The result does not change even by exchanging $\pm$ of $\mathcal{O}_\pm$ and 
$\bar{\mathcal{O}}_\pm$ in \eqref{type2}.
As in \eqref{prefactor} the total contribution is with the pre-factor $  f^2/2$. The sum of the two types of contribution reproduces \eqref{rhss} if we use 
$\langle \mathcal{T}_\lambda |\mathcal{T}_\lambda \rangle = \frac12 C_+ C_-$.

\section{Dual bulk interpretation}
\label{bulk}

In the previous section we have computed the Higgs masses of spin $s$ fields using the CFT technique.
In principle the mass can be computed directly from the bulk higher spin theory.
In fact, it was pointed out in \cite{Girardello:2002pp} that the mass term would arise from the one-loop corrections of spin $s$ propagator when we assign non-standard boundary conditions to bulk fields. 
However, it is technically difficult to extract the information of the mass from the one-loop computations.
For the simple example with $s=2$, the explicit value has been computed in \cite{Hikida:2015nfa} by following the previous works \cite{Porrati:2001db,Duff:2004wh,Aharony:2006hz}.
In principle, we can generalize their method to the case with $s > 2$, but it seems to be quite complicated.

\begin{figure}
  \centering
  \includegraphics[width=4.7cm]{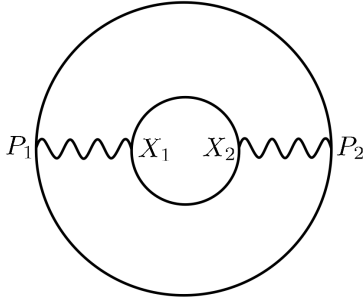}
  \caption{One-loop diagram for a current-current two point function $\langle J^{(s)}_\pm (P_1) J^{(s)}_\pm (P_2) \rangle $}
  \label{witten1}
\end{figure}

We take a different route to extract the information of mass from the one-loop effects on spin $s$ propagator.
Instead of bulk-to-bulk propagator, we consider boundary-to-boundary one, which is equivalent to the two point function of boundary spin $s$ current $\langle J^{(s)}_\pm (z) J^{(s)}_\pm (w) \rangle $. Acting $\partial_{\bar{z}} \partial_{\bar{w}}$, we can compute $|\bar \partial J^{(s)}_\pm |^2$ from the bulk higher spin theory. With \eqref{lhss} we can read off the anomalous dimension of dual spin $s$ current and the Higgs mass of the spin $s$ field can be obtained from \eqref{dictionary}. 
As mentioned above, the mass term arises from the one-loop correction  on spin $s$ propagator.
Therefore, in case of boundary-to-boundary propagator, we need to compute the one-loop diagram  as in figure \ref{witten1}.
As was pointed out, e.g., in \cite{Aharony:2006hz}, we can easily see how the corrections at the higher order of $f^2$ enter from the bulk gravity viewpoints.
Comparing the above CFT method,
we obtain the higher order corrections in the mass formula of \eqref{Msp} and \eqref{Msm}.

\subsection{One loop contributions}
\label{1loop}

In this section, we mainly work on arbitrary dimensional AdS$_{d+1}$ space.
We adopt the embedding formulation to study the higher spin theory on the space,
See appendix \ref{basics} for some details of the formulation.
We describe the space by a hypersurface  $X^2 = -1$ $(X^0 > 0)$
in $d+2$ Minkowski space. The boundary can be represented by the light rays $P^2 = 0$ and $P \sim \lambda P$ with $\lambda \in \mathbb{R}$. 
Using the coordinates, we denote the bulk-to-bulk propagator as $\Pi_{\Delta , 0} (X_1 ,X_2)$ and the bulk-to-boundary propagator as $\Pi_{\Delta , 0} (X ,P)$ for a scalar field.

In the CFT at the $d$-dimensional AdS boundary,
we introduce two complex single-trace operators $\mathcal{O}_\pm$ with scaling dimensions $\Delta_\pm$ satisfying $\Delta_+ + \Delta_- = d$.
These operators are dual to bulk complex scalars  $\phi^\pm$ with the same mass but different boundary conditions.
As in \eqref{defop2}, we consider the deformation by the double-trace operator as%
\footnote{We do not include a factor $1/2$ in \eqref{defop}. The factor $1/2$ is introduced there to cancel the Jacobian arising from the change of worldsheet coordinates as $z = \sigma_1 + i \sigma_0, \bar z = \sigma_1 - i \sigma_0$.}
\begin{align}
 \Delta S = - f \int_{\partial} d P \mathcal{T}_d (P)\, , \quad
 \mathcal{T}_d =  \mathcal{O}_+ \mathcal{O}_- + \bar{\mathcal{O}}_- \bar{\mathcal{O}}_+  \, .
 \label{def2}
\end{align}
It was shown in \cite{Witten:2001ua} that the deformation corresponds to the change of boundary condition for $\phi^\pm$. 
After the deformation the boundary conditions for the two bulk fields $\phi^\pm$ are mixed, and the effects can be removed by the rotation of the fields. Utilizing the rotation, the propagators $\langle \phi^\alpha  (X) \phi^\beta (Y) \rangle = \Pi^{\alpha \beta} (X,Y)$ $(\alpha , \beta = \pm)$ can be obtained as \cite{Aharony:2005sh,Hikida:2015nfa}
\begin{align}
 \Pi^{\alpha \beta} (X,Y) = \frac{1}{1 + \tilde f^2}
 \begin{pmatrix}  
 \Pi_{\Delta_+ ,0}  + \tilde f ^2  \Pi_{\Delta_- ,0} 
  &  \tilde f \Pi_{\Delta_- ,0} - \tilde f  \Pi_{\Delta_+ ,0} \\  
    \tilde f \Pi_{\Delta_- ,0}  - \tilde f  \Pi_{\Delta_+ ,0} 
  &   \Pi_{\Delta_- ,0} + \tilde f ^2  \Pi_{\Delta_+ ,0}  \\    \end{pmatrix}
  \label{defprop}
\end{align}
with  $\tilde f = 2 (\Delta_+ - d/2)$. 
We are interested in the case with $\Delta_\pm = (d\pm1)/2$, which lead to
 $\tilde f = f$.
 If we take $f \to \infty$, then the boundary conditions of $\phi^\pm$ are exchanged as we can see in \eqref{defprop}.

We would like to examine the loop effects of the Witten diagram in the figure \ref{witten1}.
The effects can be read off from the two point function of the bulk currents $\langle \hat J^{(s)}_\pm (x) \hat J^{(s)}_\pm (y) \rangle $.
Here the bulk currents can be obtained from the interaction term in \eqref{bulk3pt} as $\hat J^{(s)}_\pm = \hat J^{(s)}_1 \pm \hat J^{(s)}_2$, where
\begin{align}
 \hat J^{(s)}_1 = \bar \phi^+ \nabla_{\mu_1} \cdots \nabla_{\mu_s} \phi^+ \, , \quad
 \hat  J^{(s)}_2 = \bar \phi^- \nabla_{\mu_1} \cdots \nabla_{\mu_s} \phi^- \, .
\end{align}
Since the bulk currents are written in terms of bilinears of scalar fields, the two point functions can be evaluated by the product of two bulk-to-bulk propagators for the scalar fields in \eqref{defprop}.
The two point function is given by the sum over the four contributions as (see \cite{Hikida:2015nfa} for spin 2 case) 
\begin{align}
 &\langle \hat J^{(s)}_\pm (x) \hat J^{(s)}_\pm (y) \rangle  \\
&  \quad = \langle \hat J_1^{(s)} (x) \hat J_1^{(s)} (y) \rangle \pm
 \langle \hat J_1^{(s)} (x) \hat J_2^{(s)} (y) \rangle  \pm
 \langle \hat J_2^{(s)} (x) \hat J_1^{(s)} (y) \rangle + 
 \langle \hat J_2^{(s)} (x) \hat J_2^{(s)} (y) \rangle  \, . \nonumber
\end{align}
Since we know that there is no contribution to the scaling dimension from $f$ independent term, the non-trivial contribution from each term satisfies%
\begin{align} 
 \langle \hat J_1^{(s)} (x) \hat J_1^{(s)} (y) \rangle_f =
 \langle \hat J_2^{(s)} (x) \hat J_2^{(s)} (y) \rangle_f  
 = 
 -  \langle \hat J_1^{(s)} (x) \hat J_2^{(s)} (y) \rangle_f =
 -  \langle \hat J_2^{(s)} (x) \hat J_1^{(s)} (y) \rangle_f \, ,
\end{align}
which can be seen from the explicit form of the scalar propagators in \eqref{defprop}.
This expression implies that the conformal dimension for so$(3)_R$ singlet current does not receive any correction at the leading order of $1/c$. Therefore, the result in \eqref{Msp} can be easily obtained  from the viewpoint of bulk theory.
Moreover, the propagators in \eqref{defprop} implies the following important fact.
Once we have expression in the first order of $f^2$, the final result is obtained simply replacing $f^2$ by $f^2/(1+f^2)^2$. From the results in \cite{Hikida:2015nfa}, the same conclusion can be obtained for the deformations of fermionic operators.
The full order expression of $f^2$ in \eqref{Msm} can be obtained in this way.  If we take $f \to \infty$, then these masses vanish. This is consistent with the fact that the effect of the deformation at the $f \to \infty$ limit is just exchanging the  boundary conditions of $\phi^\pm$ and we know that the higher spin gauge symmetry is not broken there.

\subsection{Relation to the CFT method}
\label{relation}

As argued above, the Higgs masses should be read off from the one-loop Witten diagram for the current-current two point functions $\langle J^{(s)} _\pm (P_1) J^{(s)}_\pm (P_2) \rangle $ as in figure \ref{witten1}.
These quantities have been computed by the CFT method at the leading order of $f^2$, therefore we could relate the two ways of computation. With the relation, we can see how the higher order corrections  of $f^2$ would be computed in the CFT language.

\begin{figure}
  \centering
  \includegraphics[width=10cm]{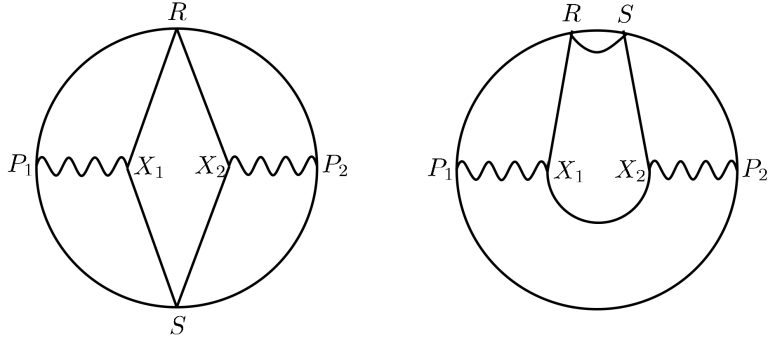}
  \caption{Contributions to anomalous dimension with the extra insertion of boundary operators}
  \label{witten2}
\end{figure}

In the one-loop diagram of figure \ref{witten1}, there are one scalar propagator along the upper solid line and another along the lower solid line. The contributions to the anomalous dimension come from the shift in propagators as in \eqref{defprop}. Let us first examine  the following term at the first order of $f$ as
\begin{align}
\Pi^{\pm \mp} (X,Y) = \tilde f (\Pi_{\Delta_- , 0} (X,Y) - \Pi_{\Delta_+ ,0} (X,Y) )\, .
\end{align}
This effect must be due to the insertion of a boundary deformation operator in \eqref{def2}.
In fact, the relation \eqref{split} implies that
\begin{align}
\label{rhsc}
   f   \int_{\partial} d R \Pi_{\Delta , 0} (X_1 , R) \Pi_{d - \Delta , 0} (X_2 , R ) 
 = f (d - 2 \Delta ) ( \Pi_{\Delta,0} (X_1 , X_2) - \Pi_{d- \Delta,0} (X_1 , X_2) ) \, .
\end{align}
Thus with $\Delta = \Delta_\pm$ the deformed propagator $\Pi^{\pm \mp} (X_1,X_2)$  in \eqref{defprop} can be written in terms of two boundary-to-bulk propagators at the first order of $f$ as expected. With this expression, this type of contribution can be written in terms of Witten diagram as in the left one of figure \ref{witten2}, 
which correspond to the product of CFT  three point functions in \eqref{type1}.

In a similar way, we can see that the right diagram of  figure \ref{witten2} can be regarded as 
 the contribution  to the propagator  $\Pi^{\pm \pm} (X_1,X_2)$   in \eqref{defprop} at the order of $f^2$. The diagram corresponds to the product of four and two point functions in \eqref{type2}.
After the two insertions of boundary operators, a propagator between $X_1$ and $X_2$ becomes
\begin{align}
\label{lhsc}
& f^2 \int d R d S \Pi_{\Delta , 0} (X_1 , R) \Pi_{d- \Delta , 0} (R , S)  \Pi_{\Delta , 0} (S , X_2) \\
 & \quad = f^2 (d - 2 \Delta )  \int d S  \Pi_{d - \Delta , 0} (X_1 , S)  \Pi_{ \Delta , 0} (S , X_2) \nonumber \\
 & \quad =  -  \tilde f^2 ( \Pi_{\Delta , 0} (X_1 , X_2)  - \Pi_{d - \Delta , 0} (X_1 , X_2)) \, , \nonumber
\end{align} 
where we have used \eqref{idelta} and \eqref{rhsc}.
It is the contribution to $\Pi^{\pm \pm} (X_1,X_2)$  in \eqref{defprop} for $\Delta = \Delta_\pm$ at the order of $f^2$ as expected.

\begin{figure}
  \centering
  \includegraphics[width=4.7cm]{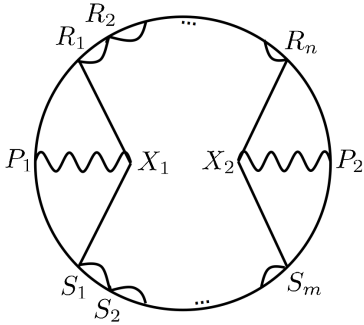}
  \caption{A contribution at the order of  $f^{n+m}$ from the insertion of   $(n+m)$ boundary operators}
  \label{witten3}
\end{figure}

From the experience with a few boundary insertions, we can guess that
the   contributions at the order of $f^{n+m}$ come from the diagrams with the insertion of $n+m$ boundary operators  as in figure \ref{witten3}. They are the Witten diagrams corresponding to the $(n+m)$-th order contributions in the conformal perturbation theory.
Let us examine how the bulk scalar propagators change due to the insertions of many boundary operators. 
Since the insertion of the boundary interaction changes the boundary scaling dimension from $\Delta$ to $d - \Delta$, the propagator  $\Pi^{\pm \mp} (X_1,X_2)$ can include odd number of insertions and the propagator $\Pi^{\pm \pm} (X_1,X_2)$ can include even number of insertions.
Using \eqref{idelta} we can see that the $2n$ number of extra boundary insertions only give factor  $ (-  \tilde f ^2 )^n $ after the integration over the insertion points. 
From the expansion of the boundary insertions  as
\begin{align}
\exp \left( f \int_{\partial} d Q \mathcal{T}_d (Q) \right) = 1 + f \int_{\partial} d Q \mathcal{T} _d (Q)
 +\frac{f^2}{2}  \int_{\partial} d Q_1 d Q_2  \mathcal{T}_d (Q_1)  \mathcal{T} _d(Q_2) + \cdots \, ,
\end{align}
we obtain the factor $1/(n!)$ for the term of the order $f^n$. However, we need to consider  all possible permutations of  the boundary points $Q_i$, which yields the factor $n!$. Thus totally, we have $1/(n!) \cdot n! = 1$ for the term of the order $f^n$. After summing over the all order contributions, the propagators with even number insertions become
\begin{align}
\Pi_{\Delta , 0} (X_1 , X_2) - 
\frac{\tilde f ^2}{1 + \tilde f^2} ( \Pi_{\Delta , 0} (X_1 , X_2) -  \Pi_{d - \Delta , 0} (X_1 , X_2)) \, , 
\end{align}
which reproduces  $\Pi_{\pm \pm} (X_1,X_2)$   in \eqref{defprop}.  
Similarly, the propagators with odd number insertions become
\begin{align}
\frac{1}{1 + \tilde f^2} (\tilde f \Pi_{\Delta_- , 0} (X_1 , X_2) - \tilde f  \Pi_{\Delta_+ , 0} (X_1 , X_2)) \, ,
\end{align}
which reproduces    $\Pi_{\pm \mp} (X_1,X_2)$   in \eqref{defprop}. In this way, we have confirmed that the full order results in \eqref{Msp} and \eqref{Msm} can be obtained also by summing over all order corrections in the conformal perturbation theory from the boundary viewpoints.

\section{Conclusion}
\label{conclusion}

In \cite{Creutzig:2014ula} a holographic duality between $\mathcal{N}=3$ coset model \eqref{coset} and a 3d Prokushkin-Vasiliev theory with extended supersymmetry \cite{Prokushkin:1998bq} was proposed. In \cite{Hikida:2015nfa} most of the higher spin symmetry was broken by adding a deformation term \eqref{def} to the CFT.
The symmetry breaking induces mass to the higher spin fields, and the explicit value of the mass was computed for a spin 2 field in \cite{Hikida:2015nfa}.
The aim of  this paper was to extend the analysis  to the case with generic spin $s > 2$.
We have computed the anomalous dimensions of higher spin currents in the boundary theory in various ways.
Using these results, we have obtained  the mass formula for higher spin fields in the bulk theory as in \eqref{Msp} for the so$(3)_R$ singlets and \eqref{Msm} for the so$(3)_R$ triplets.
The results are at the first order of $1/c$. At this order, we can utilize the free ghost system \eqref{freeghost} and the classical dual bulk theory. Combining these techniques, we can obtain expressions at the full order of the perturbation parameter $f$ in \eqref{def}.

The main motivation to study the breaking of higher spin symmetry is to understand a possible relation to superstring theory. 
There are only few candidate superstring theories. In the case with pure NSNS-backgrounds, generic arguments in \cite{Argurio:2000tg} say that only three types of target space of  superstring theory are consistent 
with the boundary $\mathcal{N}=3$ superconformal symmetry. Comparing the BPS spectrum of superstring theory in \cite{Argurio:2000xm}, it was conjectured in  \cite{Hikida:2015nfa} that the target space of the related superstring theory should be AdS$_3 \times$M$^7$ with M$^7 =$SU(3)$/$U$(1) $ or  SO$(5)/$SO$(3) $.  Since both cases lead to the same BPS spectrum, we can argue that they are related by a marginal deformation. 
Not much is  known in the case with RR-flux.

Unfortunately, it is not a straightforward task to compare the results obtained here with the string spectrum. Let us first recall the situation in \cite{Chang:2012kt}. The authors proposed a duality between a 4d extended Vasiliev theory with U($M$) CP factor and a 3d U($N$) $\times$ U($M$) Chern-Simons-matter theory known as the ABJ(M) theory \cite{Aharony:2008ug,Aharony:2008gk}. 
For the duality with classical higher spin theory, we take large $N$ since $1/N$ is related to the Newton constant $G_N$ but we keep the size of CP factor, $M$, finite.
The ABJ theory is known to be dual to superstring theory on AdS$_4 \times \mathbb{C}$P$^3$ with discrete torsion, and we can see the relation between the higher spin theory and the superstring theory by combining the above two dualities.
The amount of discrete torsion is related to the difference $N-M$, and the string (or classical supergravity) picture is relevant only for $M , N $ very large but $N-M$ finite. For the parameter region with $N \gg M$, the amount of discrete torsion is very large and the string picture is not so clear.

The coset model \eqref{coset} also has two parameters $N,M$, and their interpretation in terms of dual higher spin theory is the same as in \cite{Chang:2012kt}.
Therefore, we need to take large $N$ but finite $M$ for the higher spin holography.
In this case it is not so clear what kind of superstring theory is related to these theories, but from the analogy to the arguments in \cite{Chang:2012kt} we believe that the string picture is available only when both $M$ and $N$ are large. 
In other words, we have neglected $M/N$-corrections in this analysis, but they would be relevant for the comparison to superstring theory as argued below. For large $N$ and finite $M$, the higher spin theory is still expected to be dual to some superstring theory but at the region where physical interpretation in terms of superstring theory is difficult to obtain.

Firstly, a specific deformation \eqref{def} has been considered in this paper, but
it was found in \cite{Hikida:2015nfa} that there are other types of marginal deformations preserving $\mathcal{N}=3$ superconformal symmetry. We used the simplest one and the deformations seem to give similar effects for $N \gg M$. However,  for $N \sim M$ the effects will probably be different, and we should specify which combination of marginal deformations corresponds to turning on string tension.

Secondly, we found that the so$(3)_R$ singlet fields do not receive any corrections at the order of $1/N$ as in \eqref{Msp}.
The result is actually expected since our deformation is of the same type as the one in \cite{Gaberdiel:2013jpa}. They considered the deformation which does not break the higher spin symmetry at the leading order of $1/N$ and $f$. The deformation operator is characterized by the eigenvalue of $J^{(3)}_0$ and our deformation operator in \eqref{defop} has the required eigenvalue.  
It was also argued that the higher spin currents are no longer conserved if we consider higher order of $1/N$ and $f$. Therefore we expect that masses will be generated for $s=3,4,\ldots$ once we include the higher order effects of $1/N$.
It sounds odd that the so$(3)_R$ singlets and triplets get masses at different order of $1/N$.
However, this puzzle might be resolved if we include $M/N$-corrections.
It is natural to have non-trivial masses at the order of $1/c \cdot M/N$.
These terms become of $1/c$ order for $N \sim M$, and this is the same as those for the so$(3)_R$ triplet fields.

Thirdly, the mass formula for the so$(3)_R$ triplet fields was obtained in \eqref{Msm}, and
the formula resembles the Regge spectrum on flat space-time. 
We may compare it with the mass formula for superstrings, say, with pure NSNS-flux.
However, as argued above, there could be corrections at the order $M/N$.
For $N \sim M$ they become of the order $N^0$ and could modify the mass formula significantly.
For instance, we may find other spin dependence in the mass spectrum which is compatible to that for superstrings  with RR-flux.
As a future problem, we would like to compute the $M/N$-corrections of the mass formula in order to compare with the string spectrum.

There are other open problems and one of them is given as follows.
When we obtained the contributions at higher order of $f^2$, we have used the known results on  classical bulk theory both for bosonic and fermionic deformation operators in subsection \ref{1loop}. We have tried to interpret the contributions from the boundary viewpoints only for the bosonic ones in subsection \ref{relation}. It would be nice if we could extend the analysis for the fermionic ones, but for this purpose we need to generalize the embedding formalism also for the spinor tensor fields on AdS$_{d+1}$.

\subsection*{Acknowledgements}

We are grateful to S.~Iso, Y.~Kazama, M.~Gaberdiel  and P.~ B.~R{\o}nne for useful discussions.
The work of YH was supported in part by JSPS KAKENHI Grant Number 24740170. 
The work of TC is supported by NSERC grant number RES0019997.

\appendix

\section{Higher spin superalgebra and free ghost system}
\label{BHS}

The higher spin superalgebra shs$_{M'}[\lambda]$ introduced in \eqref{shs}
can be realized by the symmetry of the free ghost system 
\cite{Bergshoeff:1991dz} (see also \cite{Moradi:2012xd,Creutzig:2013tja})
\begin{align}
S = \frac{1}{\pi} \int d^2 z \sum_{A=1}^{M'} \left\{ \beta_A \bar \partial \gamma_A +  b_A \bar \partial c_A \right \} \, ,
\end{align}
which lead to the operator products as 
\begin{align}
 \gamma_A (z) \beta_B (w) \sim \frac{\delta_{AB}}{z-w} \, , \quad
 c_A (z) b_B (w) \sim \frac{\delta_{AB}}{z-w} \, .
\end{align}
Here $A,B = 1 , 2 , \ldots , M'$ and
the conformal weights for $b_A,c_A,\beta_A$ and $\gamma_A$ are
$(1 + \lambda)/2 , (1 - \lambda )/2 , \lambda/2 $ and $ 1 - \lambda/2$.

The truncated algebra shs$_{2}^T[1/2]$ can be generated by the free ghosts $b_A,c_A,\beta_A$ and $\gamma_A$ $(A=1,2)$ with the conformal weights $3/4,1/4,1/4$ and $3/4$, respectively.
In \cite{Bergshoeff:1991dz}, the explicit expression by free ghost system is given for higher spin currents with shs$[\lambda]$ as the wedge subalgebra, and the truncation from shs$[1/2]$ to  shs$_1^T [1/2]$ was also argued. 
Using the results we can realize 
shs$_2^T [1/2]$ as the wedge subalgebra of the one generated by
\begin{align}
\label{hs211}
 &[ V^{(s)}  (z) ]_{AB} = \sum_{i=0}^{s-1} \tilde a^i (s , 1/2) (\partial^i \beta_A) (\partial^{s-1-i} \gamma_B )
  + \sum_{i=0}^{s-1} \tilde a^i (s , 3/2) (\partial^i b_A ) (\partial^{s - 1 - i} c_B) \, , \\
 & [Q^{(s)}  (z) ]_{AB} = \sum_{i=1}^{s-1} \tilde \alpha^i (s , 1/2) (\partial^i \beta_A ) (\partial^{s-1-i} c_B ) 
 + \sum_{i=0}^{s-2} \tilde \beta^i (s , 1/2) (\partial^i b_A ) (\partial^{s-2-i} \gamma_B ) \, , 
 \nonumber
\end{align}
where $\tilde a^i (s , \lambda )$ is defined in \eqref{tildea0} and
\begin{align}
\label{tildea}
 &\tilde \alpha ^i (s , \lambda) = 2   \begin{pmatrix} s-1 \\ i \end{pmatrix}  \frac{(-1)^i}{(s)_{s-1}}
  (\lambda - s + 1)_i (2 - \lambda - s)_{s - 1 - i} \, , \\
 &\tilde \beta ^i (
 s , \lambda) =    \begin{pmatrix} s-2 \\ i \end{pmatrix}  \frac{(-1)^i}{(s)_{s-2}}
  (\lambda - s + 1)_i (2 - \lambda - s)_{s - 2 - i} \, .  \nonumber
\end{align}

The ${\cal N}=3$ superconformal algebra can be obtained as  a subalgebra
generated by the low spin currents defined in  \eqref{hs211}.
The explicit expression
of the low spin generators in \eqref{hs211}  can be given as
\begin{align}
& T = \frac{1}{4} \sum_{A=1}^2 \left[3 (\partial \beta_A) \gamma_A - \beta_A \partial \gamma_A + (\partial b_A) c_A  - 3 b_A \partial c_A \right] \, , \nonumber \\
&  J^+ = \beta_1 \gamma_2 + b_1 c_2 \, ,  \quad
   J^- = \beta_2 \gamma_1 + b_2 c_1  \, , \quad
   J^3 = \frac{1}{2} \sum_{A=1}^2 (-1)^{A-1} \left [ \beta_A \gamma_A + b_A c_A \right] \, ,  \\
 &G^+ =  ( \partial \beta_1 ) c_2 - \beta_1 \partial c_2 + 2 b_1 \gamma_2 \, , \quad
   G^- =  ( \partial \beta_2 ) c_1 - \beta_2 \partial c_1 + 2 b_2 \gamma_1 \,  , \nonumber \\
  &G^3 = \frac{1}{2} \sum_{A=1}^2 (-1)^{A-1} \left [ (\partial \beta_A ) c_A - \beta_A \partial c_A + 2 b_A \gamma_A  \right] \, , \quad  \Psi = \frac{1}{2} \sum_{A=1} ^2\beta_A c_A \, .\nonumber
\end{align}
We can check that these generators satisfy 
 the operator product expansions for the ${\cal N}=3$ superconformal algebra 
 with $c=k=0$.  See \cite{Miki:1989ri} for some details of the ${\cal N}=3$ algebra.

\section{Basics for bulk analysis}
 \label{basics}
 
In section \ref{bulk} we study the Higgs phenomenon in the viewpoint of bulk theory.
In this appendix, we introduce basic tools for it.
First we review the embedding formalism for tensor fields on AdS$_{d+1}$ or conformal fields on $d$-dimensional flat space-time,
see, for examples, \cite{Cornalba:2009ax,Weinberg:2010fx,Costa:2011mg,Costa:2011dw,SimmonsDuffin:2012uy}.
Then we give several important properties of AdS propagators, in particular, the split representation of them in \cite{Costa:2014kfa}, see also \cite{Leonhardt:2003qu,Leonhardt:2003sn}.
Finally we give integral formulas which are used in subsection \ref{relation}.

\subsection{Embedding formalism}

Euclidean AdS$_{d+1}$ space can be described by a hypersurface in $d+2$ Minkowski space defined as 
\begin{align}
 X^2 = -1 \, , \quad X^0 > 0 \, .
\end{align}
The isometry group is SO$(d+1,1)$. In the light-cone coordinates $X^A$ $(A=+,-,1,\ldots , d)$, the metric is given by
\begin{align}
 X^2 = \eta_{AB} X^A X^B = - X^+ X^- + \delta_{ab} X^a X^b 
\end{align}
with $a=1,\ldots d$. The Poincar\'e coordinates are parametrized as 
\begin{align}
X = \frac1z (1 , z^2 + y^2 , y^a) \, .
\label{poincare}
\end{align}
Near the AdS boundary, the hypersurface approaches the light-cone $X^2 = 0$. Thus the boundary can be represented by the light rays, which may be described by $P^A$ satisfying
\begin{align}
 P^2 = 0 \, , \quad P \sim \lambda P
\end{align}
with $\lambda \in \mathbb{R}$. In the Poincar\'e patch, the boundary is parametrized as
\begin{align}
P = (1 , y^2 , y^a) \, .
\label{bpoincare}
\end{align}

A totally symmetric traceless tensor $h_{\mu_1 \ldots \mu_s} (x)$ on AdS$_{d+1}$ can be described by a SO$(d+1,1)$ tensor $H_{A_1 \ldots A_s} (X)$ on the embedding space. These tensors are related as
\begin{align}
 h_{\mu_1 \ldots \mu_s} (x) = \frac{\partial X^{A_1}}{\partial x^{\mu_1}} 
  \cdots \frac{\partial X^{A_s}}{\partial x^{\mu_s}}  H_{A_1 \ldots A_s} (X) \, ,
\end{align}
which implies that
\begin{align}
 X^{A_1}H_{A_1 \ldots A_s} (X) = 0
 \label{transverse}
\end{align}
due to $X^2 = -1$. Introducing auxiliary variables $W^{A}$, we define
\begin{align}
H(X,W) = W^{A_1} \cdots W^{A_s} H_{A_1 \ldots A_s} (X) \, .
\end{align}
We assign $W^2 = 0$ from the traceless condition and $W \cdot X = 0$ from
the transverse one \eqref{transverse}.  On the boundary given by the light-cone $P^2 = 0$, we define a totally symmetric traceless tensor $F_{A_1 \ldots A_s} (P)$.  We require $F(\lambda P) = \lambda^{-\Delta} F(P) $ for $\lambda > 0$, where we denote the scaling dimension as $\Delta$. Moreover, the condition to be tangent to the light-cone $P=0$ requires $ P^{A_1}F_{A_1 \ldots A_s} (X) = 0$.
Introducing $Z^A$ we define
\begin{align}
F(P,Z) = Z^{A_1} \cdots Z^{A_s} P_{A_1 \ldots A_s} (X) \, ,
\end{align}
where we assign $Z^2 = 0$ and $Z \cdot P = 0$. The transverse condition can be encoded by requiring
$F(P ,Z + \alpha P) = F(P,Z) $ for all $\alpha$.

\subsection{AdS propagators}

We consider a  spin $s$ field with dual scaling dimension $\Delta$, which propagates from $X_1$ to $X_2$ and with polarization vectors $W_1$ and $W_2$, respectively. We represent the bulk-to-bulk propagator as $\Pi_{\Delta , s} (X_1 ,X_2 ; W_1 , W_2)$. The bulk-to-boundary propagator may be then represented as $\Pi_{\Delta , s} (X ,P ; W , Z)$. The structure of the bulk-to-boundary propagator can be fixed by conformal symmetry. 
In \cite{Costa:2014kfa}, it was claimed that the bulk-to-bulk propagator can be written in terms of 
 the following AdS harmonic function as%
 \footnote{We changed the normalization of boundary-to-bulk operator by $- 4 \nu^2$.}
\begin{align}
& \Omega_{\nu , s} (X_1 , X_2 ; W_1 , W_2) \\
 & \quad  = \frac{1}{16 \pi s! (\frac{d}{2} - 1)_s \nu^2} 
 \int_{\partial} dP \Pi_{\frac{d}{2} + i \nu , s} (X_1 , P ; W_1 . D_Z) \Pi_{\frac{d}{2} - i \nu ,s} (X_2 , P ; W_2 , Z) \, .  \nonumber
\end{align} 
Here $(a)_n$ is defined in \eqref{Poch} and the operator $D_Z$ is given by
\begin{align}
 D_Z^A = \left(\frac{d}{2} - 1 + Z \cdot \frac{\partial}{\partial Z}\right) \frac{\partial}{\partial Z_A} - \frac{1}{2} Z^A \frac{\partial^2}{\partial Z \cdot \partial Z} \, .
\end{align}
The harmonic function is found to satisfy \cite{Leonhardt:2003qu,Costa:2014kfa}
\begin{align}
\label{split}
& \Omega_{\nu , s} (X_1 , X_2 ; W_1 , W_2)  \\
& \quad  = \frac{ 1}{ 8 i \pi \nu }
 \left( \Pi_{\frac{d}{2} + i \nu , s} (X_1 , X_2 ; W_1,W_2) -  \Pi_{\frac{d}{2} - i \nu , s} (X_1 , X_2 ; W_1,W_2)\right) \, . \nonumber
\end{align}

\begin{figure}
  \centering
  \includegraphics[width=4.7cm]{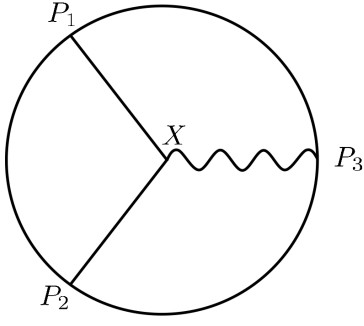}
  \caption{Witten diagram for a three point function}
  \label{3ptfn}
\end{figure}

In the CFT method we have used the three point functions in \eqref{3pt}. With the above propagators we can express them as Witten diagrams in terms of bulk theory as in figure \ref{3ptfn}.
 We introduce a spin $s$ field $h_{\mu_1 \ldots \mu_s}$ and a complex scalar $\phi$, $\bar \phi$.  As argued in \cite{Costa:2014kfa} the generic three point coupling may be set in the form of 
 \begin{align}
 \label{bulk3pt}
 g \int_{\text{AdS}} dx \sqrt{g} (\bar \phi \nabla_{\mu_1} \cdots \nabla_{\mu_s} \phi) h^{\mu_1 \ldots \mu_s} \, .
 \end{align}
 Here we have assumed the transverse condition $\nabla^{\mu_1}h_{\mu_1 \ldots \mu_s} = 0 $.
 Let us denote $\Delta_0 , \Delta_s$ as the dual scaling dimensions for the complex scalar  and the spin $s$ field. 
Then the Witten diagram can be evaluated as
\begin{align} 
&\langle \mathcal{O}_{\Delta_0} (P_1) \bar{\mathcal{O}}_{\Delta_0} (P_2) J^{(s)} (P_3,Z)  \rangle  \\
& \quad =
 g \int_{\text{AdS}} d X  \Pi_{\Delta_0 , 0} (P_1 ,X) \frac{\Pi_{\Delta_s , s} (X, P_3 , K , Z  ) ( W \cdot \nabla)^J \Pi_{\Delta_0 , 0} (X, P_2)}{s! ((d-1)/2)_s} \, .
 \nonumber
\end{align}
Here $K$ is a projector operator, whose expression may be found in \cite{Costa:2014kfa} as
\begin{align}
 K_A &= \frac{d-1}{2} \left[ \frac{\partial}{\partial W^A} + X_A \left( X \cdot \frac{\partial}{\partial W}\right) \right]+ \left( W \cdot \frac{\partial}{\partial W} \right)
 \frac{\partial}{\partial W^A} \\  & \qquad \qquad +
 X_A \left( W \cdot \frac{\partial}{\partial W} \right) \left( X \cdot \frac{\partial}{\partial W} \right)- \frac12 W_A \left[ \frac{\partial^2}{\partial W \cdot \partial W}  + \left( X \cdot \frac{\partial}{\partial W} \right)^2\right] \, . \nonumber
\end{align}

\subsection{Integral formulas}
\label{IF}

The expression for the bulk-to-boundary propagators in the embedding formulation may be found in \cite{Costa:2014kfa}. For a spin 0 field, it is given by
\begin{align}
 \Pi_{\Delta,0} (X , P) = \mathcal{C}_\Delta \frac{1}{( - 2 P \cdot X)^\Delta} \, , \quad
  \mathcal{C}_\Delta  = \frac{(2 \Delta - d) \Gamma (\Delta)}{\pi^{d/2} \Gamma (\Delta - d/2)} \, .
\end{align}
The boundary-to-boundary propagator is then obtained by replacing $X$ by $Q$ with $Q^2 = 0$. 

In the main context, we need to evaluate the following integral as
\begin{align}
 I_\Delta (X,Q) &= \int_{\partial} d P \Pi_{d - \Delta ,0} (X , P) \Pi_{\Delta ,0} (Q , P)  \\
  &=  \mathcal{C}_{d - \Delta}  \mathcal{C}_\Delta  \int_{\partial} d P \frac{1}{( - 2 P \cdot X)^{d-\Delta}} \frac{1}{( - 2 P \cdot Q)^\Delta} \, . \nonumber
\end{align}
Using the Feynman parametrization
\begin{align}
 \frac{1}{\prod_i A^{a_i}_i }
  = \frac{\Gamma (\sum_i a_i)}{\prod_i \Gamma (a_i)} \int_0^\infty \prod_{i=2}^n  d q_i q_i^{a_i - 1}
  \frac{1}{( A_1 + \sum_{i=2}^n q_i A_i  ) ^{\sum_i a_i }} \, ,  
\end{align}
the integral can be written as
\begin{align}
 I_\Delta (X,Q) = \mathcal{C}_{d - \Delta}  \mathcal{C}_\Delta  \frac{\Gamma (d)}{\Gamma(d - \Delta) \Gamma (\Delta)} \int_{\partial} d P \int_0^\infty dq \frac{q^{\Delta - 1}}{( - 2 P \cdot Y)^d} 
\end{align}
with $Y = X + q Q$. In \cite{SimmonsDuffin:2012uy} a useful integral formula  was derived as
\begin{align}
 \int_{\partial} d P \frac{1}{ ( -2 P \cdot Y )^d} = \frac{\pi^{d/2} \Gamma (d/2)}{\Gamma (d)} \frac{1}{(- Y^2)^{d/2}}
\end{align}
with $Y^2 < 0$. Applying the formula we have 
\begin{align}
\nonumber
 I_\Delta (X,Q) &= \mathcal{C}_{d - \Delta}  \mathcal{C}_\Delta  \frac{\pi^{d/2} \Gamma (d/2)}{\Gamma(d - \Delta) \Gamma (\Delta)} \int_0^\infty dq \frac{q^{\Delta - 1}}{( 1 - 2 q Q \cdot X )^{d/2}}  \\ &= (d - 2 \Delta) \Pi_{\Delta, 0} (X , Q)\, .
 \label{idelta}
 \end{align}
In the last equality we have used 
\begin{align}
 \int_0^\infty dq q^{\Delta - 1} (1 + q)^{-d/2} = \frac{\Gamma (d/2 - \Delta) \Gamma (\Delta)}{\Gamma (d/2)} \, .
\end{align}


\providecommand{\href}[2]{#2}\begingroup\raggedright\endgroup

\end{document}